\documentclass[8.5pt,twoside,twocolumn]{article}
\usepackage[super,sort&compress,comma]{natbib} 
\usepackage{mhchem}
\usepackage{times,mathptmx}
\usepackage{sectsty}
\usepackage{amsmath}
\usepackage{balance} 
\usepackage{morefloats} 
\usepackage{graphicx} 
\usepackage{lastpage}
\usepackage[format=plain,justification=raggedright,singlelinecheck=false,font=small,labelfont=bf,labelsep=space]{caption} 
\usepackage{fancyhdr}
\usepackage{hyperref}
\begin{document}

\thispagestyle{plain}
\fancypagestyle{plain}{
\renewcommand{\headrulewidth}{1pt}}
\renewcommand{\thefootnote}{\fnsymbol{footnote}}
\renewcommand\footnoterule{\vspace*{1pt}%
\hrule width 3.4in height 0.4pt \vspace*{5pt}} 
\setcounter{secnumdepth}{5}

\makeatletter 
\def\subsubsection{\@startsection{subsubsection}{3}{10pt}{-1.25ex plus -1ex minus -.1ex}{0ex plus 0ex}{\normalsize\bf}} 
\def\paragraph{\@startsection{paragraph}{4}{10pt}{-1.25ex plus -1ex minus -.1ex}{0ex plus 0ex}{\normalsize\textit}} 
\renewcommand\@biblabel[1]{#1}            
\renewcommand\@makefntext[1]%
{\noindent\makebox[0pt][r]{\@thefnmark\,}#1}
\makeatother 
\renewcommand{\figurename}{\small{Fig.}~}
\sectionfont{\large}
\subsectionfont{\normalsize} 

\fancyfoot{}
\fancyhead{}
\renewcommand{\headrulewidth}{1pt} 
\renewcommand{\footrulewidth}{1pt}
\setlength{\arrayrulewidth}{1pt}
\setlength{\columnsep}{6.5mm}
\setlength\bibsep{1pt}
 
\twocolumn[
  \begin{@twocolumnfalse}
  \noindent\LARGE{\textbf{A general forcefield for accurate phonon properties of metal-organic frameworks}}
\vspace{0.6cm}

\noindent\large{\textbf{Jessica K. Bristow,\textit{$^{a}$} Jonathan M. Skelton,\textit{$^{a}$} Katrine L. Svane,\textit{$^{a}$} Aron Walsh,\textit{$^{\ast}$}{$^{a,b}$} and Julian D. Gale\textit{$^{\ast}$}{$^{c}$}}}

\vspace{0.5cm}

\noindent\textit{\small{\textbf{$^{a}$~Centre for Sustainable Chemical Technologies and Department of Chemistry, University of Bath, Claverton Down, Bath, BA2 7AY, UK. Tel: 01225 385432; E-mail: a.walsh@bath.ac.uk}, \textbf{$^{b}$~Global E$^3$ Institute and Department of Materials Science and Engineering, Yonsei University, Seoul 120-749, Korea}, \textbf{$^{c}$~Curtin Institute for Computation, Department of Chemistry, Curtin University, PO Box U1987, Perth, WA 6845, Australia. Email: j.gale@curtin.edu.au}}}


\vspace{0.6cm}

\noindent \normalsize{
We report the development of a forcefield capable of reproducing accurate lattice dynamics of metal-organic frameworks. Phonon spectra, thermodynamic and mechanical properties, such as free energies, heat capacities and bulk moduli, are calculated using the quasi-harmonic approximation to account for anharmonic behaviour due to thermal expansion. Comparison to density functional theory calculations of properties such as Gr$\mathrm{\ddot{u}}$neisen parameters, bulk moduli and thermal expansion supports the accuracy of the derived forcefield model. Material properties are also reported in a full analysis of the lattice dynamics of an initial subset of structures including: MOF-5, IRMOF-10, UiO-66, UiO-67, NOTT-300, MIL-125, MOF-74 and MOF-650.
}
\vspace{0.5cm}
 \end{@twocolumnfalse}

]

\section{Introduction}

Metal-organic frameworks (MOFs), formed from metal cations and anionic organic molecules, are versatile materials with a range of functional properties. There are a vast number of possible candidate structures from known metals and ligands, which may self-assemble into a variety of 3D MOF structures. Developing a cheap, transferable and accurate method for initial property screening of potential MOF candidates for applications such as gas absorption, explosives detection and use in solar energy conversion is therefore desirable to reduce the cost and time of experimental work. Given the tractable, but still computationally expensive, nature of Density Functional Theory (DFT) and higher levels of first principles theory, the development of forcefields capable of reproducing structural, mechanical and vibrational properties of MOFs would be highly advantageous to the large community of computational chemists interested in the thermoelastic properties of MOFs. 

A current dilemma for the development of forcefields for MOFs is the choice between transferability and accuracy. Large scale screening procedures offer a powerful tool for guiding experimental work but often involve multiple approximations, creating a level of uncertainty when used to calculate complex properties. \textit{Ab initio} derived forcefields for prediction of charges and force constants offer accurate reproduction of the properties of individual or families of MOFs with similar topologies, but lack transferability. However, one must also consider that the purpose of using a forcefield for materials with large primitive unit cells is to remove the need to conduct expensive higher level calculations in the first place. A final consideration is associated with the diversity of MOFs. Due to differing compositions and topologies, the nature of the interaction between metal and ligand ranges between ionic and covalent. To derive one forcefield capable of reproducing all such interactions is a challenging task. 

A vast number of forcefields for MOFs have already been reported; here we give a summary of the most prevalent and recent developments as a brief but broad overview. When parametrising a forcefield for MOFs there are many different approaches one can take.
Firstly, existing transferable forcefields, such as CHARMM\cite{brooks2009charmm,brooks1983charmm}, MM3\cite{allinger1989molecular,allinger1990molecular}, GAFF\cite{wang2004development} and UFF\cite{rappe1992uff}, which have been extensively derived for common organic and, to a lesser extent, inorganic compounds, offer an abundant source of reasonable parameters for the individual components of MOFs. It is therefore only the interaction between metal and ligand for which additional potential parameters must be derived. UFF4MOF\cite{addicoat2014extension} is an example of such a forcefield, reported by Addicoat \textit{et al.}, where the forcefield is an extension to the Universal forcefield (UFF). The UFF consists of multiple parameters capable of adequately reproducing the structures of organic molecules and inorganic clusters with little specific fitting. Indeed its incorporation in many user-friendly visualisation programs has increased the popularity of the UFF forcefield over many other more specifically parameterised transferable forcefields such as CHARMM. UFF4MOF employs additional atom types with parameterised valence coordination, equilibrium bond distances, effective charges and bond angles of the UFF to achieve a more accurate reproduction of the structure of clusters and periodic models of different MOFs.
A more specific modification of UFF to reproduce the interaction of IRMOF-10 with \ce{CO2} was reported
by Borycz \textit{et al}.\cite{borycz2016co2}

 Another approach is to derive bonding force constants and partial charges for each individual MOF using DFT calculations. MOF-FF is an example of a DFT-derived forcefield and was developed by Schmid \textit{et al.}\cite{bureekaew2013mof} MOF-FF is capable of reproducing the structure and properties of many MOF topologies with initial parametrisation using the MM3 forcefield. The forcefield is then further fitted for individual MOFs based on data obtained from first principles calculations. The transferability for many families of MOFs has therefore not been extensively tested. Quick-FF was published by Van Speybroeck \textit{et al.}\cite{vanduyfhuys2015quickff} and offers a method for rapid application of a forcefield to a MOF based on force constants extracted from a first principles Hessian. Quick-FF is still based on the MM3 non-bonding functional form but bonding parameters must be input by the user for the MOF of interest. The application of Quick-FF to MOF-5 and MIL-53 was reported, and was shown to reproduce both the structural parameters and the breathing behaviour of MIL-53.\cite{vanduyfhuys2015quickff} Here derivation of charges and bonding parameters remains dependent on first principles calculations. BTW-FF, our own previously reported MOF forcefield, which has been shown to reproduce the structure and properties of many different MOF topologies, also used partial charges obtained from DFT, using a Bader analysis of the charge density of each MOF considered.\cite{bristow2014transferable} 

Finally, large scale screening procedures can be based on a primitive mathematical description of the bonding in a framework or using a transferable forcefield, such as UFF, for all structures. 
High-throughput computational screening offers a valuable method for an approximate initial analysis of properties such as volumetric gas uptake or surface area. Screening also offers a means of structure prediction for hypothetical frameworks based on possible bonding considerations.\cite{wilmer2012large,yazaydin2009screening,haranczyk2015mathematical,parkes2013screening}   


Comparisons of forcefields for simple MOFs, such as MOF-5, are now standard practice and offer little evaluation of the difficultly in derivation of the forcefield. Furthermore, the transferability of parameters across variations in the topology and crystallinity are rarely extensively tested. In particular, a recent focus is on defects and disorder in MOFs. Alterations of the frameworks must still render accurate and reliable results with the same forcefield pardepenameters. 

We highlight the calculation of phonon properties to be a relatively sparsely populated area of study for MOF forcefields; indeed, many forcefields, including previous work of ours, have merely compared vibrational $\Gamma$-point frequencies and plotted IR spectra. Phonon properties are critical for the analysis of dynamic stability, particularly if soft-modes are present, energetic stability (\textit{via} free energies) and finite-temperature properties. Calculating phonon properties, such as phonon dispersion curves, with DFT is costly, and is only affordable and practical for specific MOFs of interest. Routine screening of phonon properties for large numbers of MOFs is currently only affordable with forcefield methods. 
One must also remember that experimental structures are often determined using X-ray diffraction. This can lead to inaccurate structure refinement of hydrogen positions and assignment of space groups that represent average structures. For some MOFs this may lead to loss of information regarding subtle distortions, such as non-planarity of carboxylate groups. Analysing the phonon stability of the structure, particularly for DFT calculations, is expensive and often avoided during electronic structure analysis. A transferable forcefield, not specifically fitted for each individual MOF, may allow small distortions of a framework to be identified, though the extent to which this is possible may depend on whether polarisability is included in the model or not. Optimising MOFs with a forcefield prior to using DFT may therefore be beneficial. 

A final note regarding the importance of phonon property calculations is that MOF forcefields are often fitted at 0 K to an experimental structure determined at room temperature. Furthermore, temperature dependent properties (such as bulk moduli) from a 0 K optimisation are also often compared to room temperature experiments. Incorporating the consideration of temperature through free energy minimisation is a desirable alternative solution to having to make such approximate comparisons.

Here we present a new forcefield derived with the intention to bridge the gap between accuracy and transferability, while also incorporating an extensive analysis of phonon properties. The forcefield, named \textsc{VMOF} (Vibrational Metal-Organic Framework), is derived with the intention to be transferable, accessible and accurate when reproducing the structure and dynamical properties of MOFs. \textsc{VMOF} is a development of our previously reported BTW-FF forcefield for MOFs. In this paper we report the foundations and derivation of \textsc{VMOF}, along with a comparison of initial structure parameters and mechanical properties calculated for a range of MOFs. The main focus of the paper is then on discussing the accuracy of the forcefield for reproducing phonon properties obtained from DFT. We report densities of states (DOS), infra-red (IR) spectra and temperature dependent thermodynamic properties, such as free energies, vibrational entropy, and constant volume heat capacities. We show this new forcefield to be capable of accurately reproducing properties for an initial subset of MOFs. Finally, we perform quasi-harmonic calculations that, to the best of our knowledge, have not been previously reported for the given structures, and report additonal temperature dependent structural properties.

\section{Methodology}

\subsection{First principles reference calculations}

Reference quantum mechanical calculations were conducted to have a standard method of validation of the new forcefield. The Vienna \textit{ab initio} Simulation Package (VASP)\cite{kresse1996software} code was used to perform Kohn-Sham density functional theory (DFT) calculations using the PBEsol exchange-correlation functional.\cite{perdew2008restoring} The projector-augmented wave method\cite{blochl1994projector} was used for the interaction between core and valence electrons of all atoms in the system. During optimisation, all forces were converged to values of less than 0.001 eV/$\mathrm{\AA}$ with a plane wave basis sets cut-off of 600 eV. $\Gamma$-point sampling of the Brillouin zone was considered sufficient for the MOFs owing to the unit cell dimensions of the systems, excluding MOF-74, which required a 4 $\times$ 4 $\times$ 4 k-point grid. A D3 van der Waals correction\cite{grimme2010consistent} was included and found to be necessary to remove phonon instabilities for some MOF structures. Reference calculations for the binary oxides: ZnO, Al$_{2}$O$_{3}$, TiO$_{2}$ and ZrO$_{2}$, were performed with the same convergence criteria in VASP, with the chosen polymorphs being wurtzite, corundum, rutile and baddelite, respectively. The \textbf{k}-point grid and plane wave cut-off were converged separately for each metal oxide, with final values being given in the SI along with the optimised unit cell parameters.  


\subsection{Forcefield calculations}

\textsc{VMOF} was derived using the General Utility Lattice Program (GULP) code, which has extensive capabilities suited to both inorganic and organic materials.\cite{gale1997gulp,gale2003general} \textsc{VMOF} considers the metal node and organic component as essentially separate entities interacting only by modified MM3 Buckingham potentials, in addition to the Coulomb terms. 

\begin{equation}
E_{ij}^{MM3} = \epsilon_{ij} \bigg[A\mathrm{exp}\bigg(-B\frac{d_{ij}}{d^0_{ij}}\bigg) - C\bigg(\frac{d^0_{ij}}{d_{ij}}\bigg)^6\bigg]
\label{buck}
\end{equation}

The MM3 Buckingham functional (Equation \ref{buck}) form consists of defined constants, \textit{A} (1.84 $\times$ 10$^{5}$), \textit{B} (12) and \textit{C} (2.25), and was proposed as a ``softer'' energy function to that used in MM2. MM3 has been shown to accurately reproduce hydrogen and carbon positions in many aromatic compounds.\cite{allinger1989molecular,allinger1990molecular} 
The two remaining parameters per atom type, epsilon and the van der Waals radius, were fitted to reproduce phonon stable metal oxide structures by deriving these parameters for both metal and the inorganic oxygen. The reference to inorganic oxygen here describes oxygen atoms that are coordinated only to metal atoms. Epsilon and van der Waals terms for the carboxylate oxygen and hydroxyl oxygen atom types were fitted in a relaxed fitting procedure to reproduce structural and mechanical properties of all the MOFs being tested. A further feature is the use of combination rules, where; $\epsilon_{ij}$=$\sqrt{\epsilon_{ii}\epsilon_{jj}}$ and $d_{ij}$=$\sqrt{d_{ii}d_{jj}}$, to reduce the number of parameters that require fitting, thus increasing the transferability. The long-range cut-off of the MM3 Buckingham potentials was set to 12 $\mathrm{\AA}$.

The derivation of parameters for the organic ligands was considered separately. Intramolecular bonding parameters for the ligands are taken directly from the CHARMM library and are implemented as harmonic functions. We consider the intramolecular bonding parameters between neighbouring atoms, angles between three connected neighbours and torsions between four connected atoms. A small modification was made to the CHARMM parameters for the 4-body torsion across the carboxylate head of all ligands considered, the derivation of which will be discussed later.

The total energy ($U$) can be written as; 
\begin{equation*}
 \begin{aligned}
U = \sum_{bonds} \frac{1}{2}k_r(r-r_0)^2 + \sum_{angles} \frac{1}{2}k_{\theta}(\theta - \theta_0)^2 + \\
  \sum_{dihedrals} \frac{1}{2}k_{\Psi}[1 + cos(n\Psi + \Psi_0)] + \frac{1}{2}\sum_i \sum_j \frac{q_iq_je^2}{4\pi \epsilon_0 r_{ij}} 
   \end{aligned}
\end{equation*}
where, k$_r$, k$_\theta$ and k$_\Psi$ are interatomic force constants, $r$ the distance between pairs of atoms, $\theta$ and $\Psi$ are angles, $q$ represents point charges and $\epsilon_{0}$ is the vacuum permittivity. Note that the harmonic bonding terms in GULP possess a multiplication factor of $\frac{1}{2}$, and so CHARMM force constants were appropriately scaled.

The charges of the ligands are derived within GULP using the charge equilibration model of Gasteiger\cite{gasteiger1980iterative,gasteiger1978new}, while formal charges are used for the metal nodes and inorganic oxygen atoms. Gasteiger charges were selected since, the charges are geometry independent and depend only on connectivity. Whilst other charge equilibration schemes suffer from charge delocalisation errors, Gasteiger charges do not.\cite{gasteiger1978new} Initial charge parametrisation involved taking the average charge of each atom type in a subset of common MOF ligands including; 1,4 - dicarboxylate (BDC), 1,3,5 - tricarboxylate (BTC), 4,4'-biphenyl dicarboxylate (BPDC), 2,6 - azulenedicarboxylate (AZ), 4,4' - biphenyl tricarboxylate (TPDC) and 2,7 - pyrene-dicarboxylate (PDC). Once derived for a specific atom type, the same charges are used for all the structures modelled.

\subsection{Property calculations}

\subsubsection{Mechanical properties. }

Bulk moduli (B$_{0}$) were calculated from the relevant components of the elastic constant and compliance tensors, which were determined from the analytical second derivatives of the energy with respect to strain on the system. The elastic compliance tensor is just the inverse of the elastic constant tensor. 
The reported bulk moduli calculated with the forcefield follow the Hill convention, \textit{i.e.} they are the averages of the Reuss and Voigt definitions. First principles bulk moduli for each structure were calculated at 0 K by fitting a third-order Birch-Murnaghan equation of state to energy-volume data calculated at a series of expansions and contractions about the equilibrium structure. For each expansion and contraction the atomic positions were optimised.



\subsubsection{Vibrational properties. }

IR frequencies and intensities, as well as densities of states, were post-processed from both forcefield and first principles methods to apply a consistent broadening factor of 10 cm$^{-1}$ with a frequency sampling resolution of 1 cm$^{-1}$. 
First principles calculations of dynamical matrices were performed using density functional perturbation theory within VASP to obtain $\mathrm{\epsilon^{\infty}}$ and Phonopy was used to calculate the eigenvalues. IR frequencies and intensities were then plotted with identical resolution and broadening factors.

\subsubsection{Lattice dynamics and thermodynamic properties. }

To calculate the thermodynamic properties, such as Helmholtz and Gibbs free energies, vibrational entropy and volumetric heat capacity, and the dependence with temperature, we use Phonopy.\cite{togo2015first,togo2010first} Phonopy is a python package for setting up post-processing finite-displacement phonon calculations that can be integrated with multiple first principles codes, and includes an extension for the quasi-harmonic approximation. Further details are given in the SI.

The quasi-harmonic approximation\cite{mopsik1973quasi} allows a greater number of thermodynamic properties to be computed along with their temperature-dependence. The practical reality of the quasi-harmonic approximation is to minimise the internal energy at constant volume at a given number of lattice expansions and contractions away from the global minimum structure. An equation of state is then fitted across the calculated temperature-dependent Helmholtz free energies, which changes the minimum free energy volume according to a defined temperature. The temperature dependence of the phonon frequencies is then expressed in terms of Gibbs free energy. The theory is still dependent on calculations conducted under the harmonic approximation, but the consequent volume dependence of the vibrational frequencies is an anharmonic effect. In general it is possible to use the analytical derivatives of the free energy with respect to strain to optimise all cell parameters independently within the ZSISA\cite{allan1996zero} (Zero Static Internal Stress Approximation) approximation. While this can be routinely achieved using some forcefield implementations\cite{gale1998analytical}, the requirement to compute third-order derivatives makes this particular demanding with quantum mechanical approaches.    

Throughout the calculation of thermodynamic properties a consistent Brillioun-zone phonon sampling (q-point) mesh of 32 $\times$ 32 $\times$ 32, with a symmetry tolerance of 10$^{-3}$ $\mathrm{\AA}$ for determining the space group symmetry during atomic displacement generation. 

To remove imaginary modes, observed only during compression in the DFT calculations, the corresponding Hessian eigenvectors were used to displace the atoms, thereby lowering the symmetry, and the structure relaxed accordingly. Removing imaginary modes with the forcefield required the use of a rational function optimisation approach to ensure that the Hessian has the correct final structure after optimisation.

\section{Results}

\subsection{Forcefield fitting}

\subsubsection{Universal forcefield and metal oxide parameters. }

When considering the initial derivation of parameters for each metal, besides fitting the epsilon and van der Waals radius for each species, we have also examined the influence of changing the Universal MM3 constants. This was found to be beneficial to the overall quality of the results. After extensive testing, we found that the modification of the MM3 constants by changing the B parameter to 11.5 and C parameter to 2.55, reproduced DFT and experimental structural and mechanical properties of the MOFs and metal oxides more accurately. In support of the modification of the MM3 constants, we report structural and mechanical properties for the binary oxides with the original MM3 constants (see SI) and modified constants (Table \ref{MOtable1}) in comparison with experimental values. Using the original MM3 constants results in large errors for second order elastic properties such as the elastic constants and bulk moduli.

\begin{table*}[h]
\caption{Comparison of structural and mechanical properties of metal oxides between experiment (Exp.) and the \textsc{VMOF} forcefield with modified MM3 parameters (\textsc{VMOF}). Given are the elastic constants (C$_{ij}$), bulk moduli (B$_{0}$), unit cell parameters and metal-oxygen (M-O) bond lengths. Percentage errors in the cell parameters are given in parenthesis.}
\centering
\begin{tabular}{c| c c |c c| c c |c c} 
\hline
 & \multicolumn{2}{c}{ZnO} & \multicolumn{2}{c}{ZrO$_{2}$} & \multicolumn{2}{c}{TiO$_{2}$} & \multicolumn{2}{c}{Al$_{2}$O$_{3}$} \\
Property & Exp\cite{ahuja1998elastic} & \textsc{VMOF}  & Exp\cite{fadda2009first} & \textsc{VMOF}  & Exp\cite{jian2008mechanical}  & \textsc{VMOF} & Exp\cite{gladden2004reconciliation}  & \textsc{VMOF}   \\
\hline
\textit{a} ($\mathrm{\AA}$) & 3.250  &  3.219 (0.95) & 5.070 &  5.123 (1.05)&  4.594 &  4.414 (3.92) & 4.764 & 4.870 (2.23)  \\
\textit{b} ($\mathrm{\AA}$) & 3.250 &  3.219 (0.95) & 5.070 &  5.123 (1.05)& 4.594  &  4.414 (3.92) & 4.764  & 4.870 (2.23)\\
\textit{c} ($\mathrm{\AA}$) & 5.207 &  5.014 (3.71)  & 5.070 &  5.123 (1.05) &  2.959 & 3.168 (7.06)& 13.001& 12.899 (0.78)\\
M-O ($\mathrm{\AA}$) & 1.992 &  1.967 & 2.195 &  2.218 & 1.980, 1.949 &   1.980, 1.934 &1.858 & 1.848 \\
& & & & & &  \\
C$_{11}$ (GPa) & 209.6 & 242.6 & 533.5 & 630.8 & 366.0 & 362.4 &497.3 & 564.3 \\
C$_{12}$ (GPa) & 121.1 &  108.2 & 97.86 &  131.2 &  225.0 &  337.2 &162.8 &  224.3 \\
C$_{13}$ (GPa) & 105.1 &  100.7 & - & - & - & - & 116.0 & 152.2 \\
C$_{33}$ (GPa) & 210.9 &  199.2 & - & - &- & -& 500.9 & 463.3 \\
C$_{44}$ (GPa) & - &  - & 64.26 &  125.7 & 189.0 &  213.7 &146.8 &123.9 \\
C$_{55}$ (GPa)  & 42.5 &  78.4 & - & - &  - & - &-  & -   \\
B$_{0}$ (GPa) & 183.0 &  143.2 & 243.7 &  297.7 & 282.0 &  335.1 &240.0 & 291.3\\
\hline
\end{tabular}
\label{MOtable1}
\end{table*}

The forcefield model that we have chosen to adopt to increase transferability involves formal charges at the metal node. The unit cell parameters and elastic constants of the metal oxides reproduced by the forcefield are generally reasonable (Table \ref{MOtable1}), though with a tendency to overestimate the hardness of materials. Formal charge models usually include the shell model for polarisation of the oxide ions, which can effectively soften the mechanical properties, but we maintain the use of a rigid ion model for consistency with the CHARMM parameters for the organic ligands.

\subsubsection{Ligand parameters. }
 
Force constants for the ligand parameters are taken directly from the CHARMM library. During the derivation of forcefield parameters for the metal nodes, there was one 4-body interaction with the ligand that showed a particular propensity for producing phonon unstable structures if varied. The torsion across the head of the carboxylate groups did not always remain planar during optimisation, and enforcing planarity by increasing the force constant across this bonding connection often rendered structures phonon unstable. We therefore calculated the force constant across this interaction in an isolated BDC$^{2-}$ ligand in the gas phase. The PBE0 functional\cite{adamo1999toward} was used in the NWChem program\cite{valiev2010nwchem} with the Dunning correlation consistent cc-pVTZ basis sets\cite{dunning1989gaussian} to fully relax the ligand with a 0 and 90$^{\circ}$ torsion of the carboxylate heads in relation to the aromatic ring (depicted in the SI). We calculate the energy difference between the two configurations to be 0.530 eV. To maintain transferability of the forcefield we assume little variation of this energy would occur across different aromatic dicarboxylate ligands. Therefore, it is this value that the force constant is fitted to for the torsion between O$_{carb}$-C$_{carb}$-C$_{benz}$-C$_{benz}$ for all structures with these atom types.

\subsection{Structural properties}

\begin{figure}
\centering
\includegraphics[width=8.6cm]{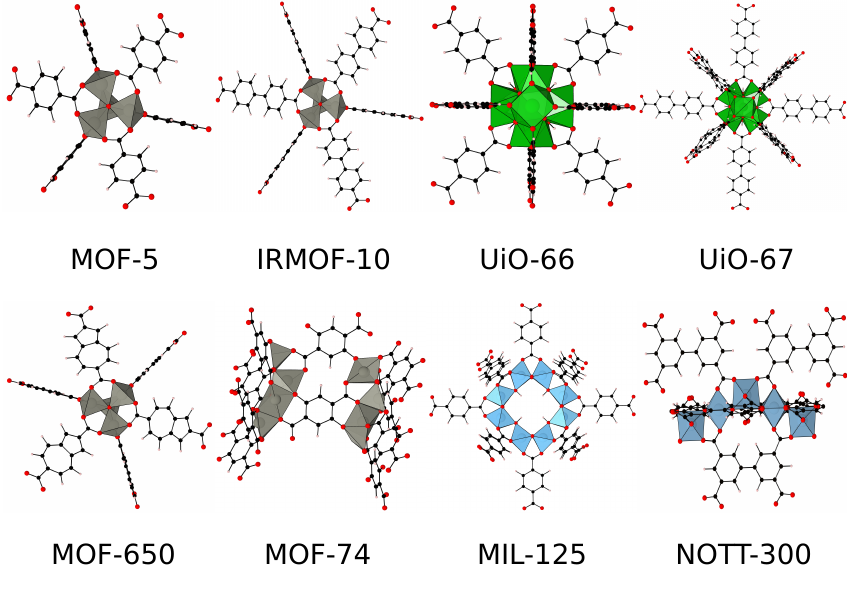}
\caption{Structures of MOF-5, IRMOF-10, UiO-66, UiO-67, MOF-650, MOF-74, MIL-125 and NOTT-300. Metal polyhedra are highlighted for Zn (silver), Zr (green), Ti (blue) and Al (grey), with atoms coloured black for carbon, white for hydrogen and red for oxygen. Compositions and symmetries are given in Table \ref{MOFparamset}.}
\label{imagesmof}
\end{figure}

Following the derivation of the forcefield parameters, mechanical and vibrational properties have been calculated for eight different MOFs (Figure \ref{imagesmof}), representing a range of ligand and metal node types: MOF-5, IRMOF-10, UIO-66, UIO-67, MOF-650, MIL-125 and NOTT-300. 

\begin{figure}
\centering
\includegraphics[width=8.6cm]{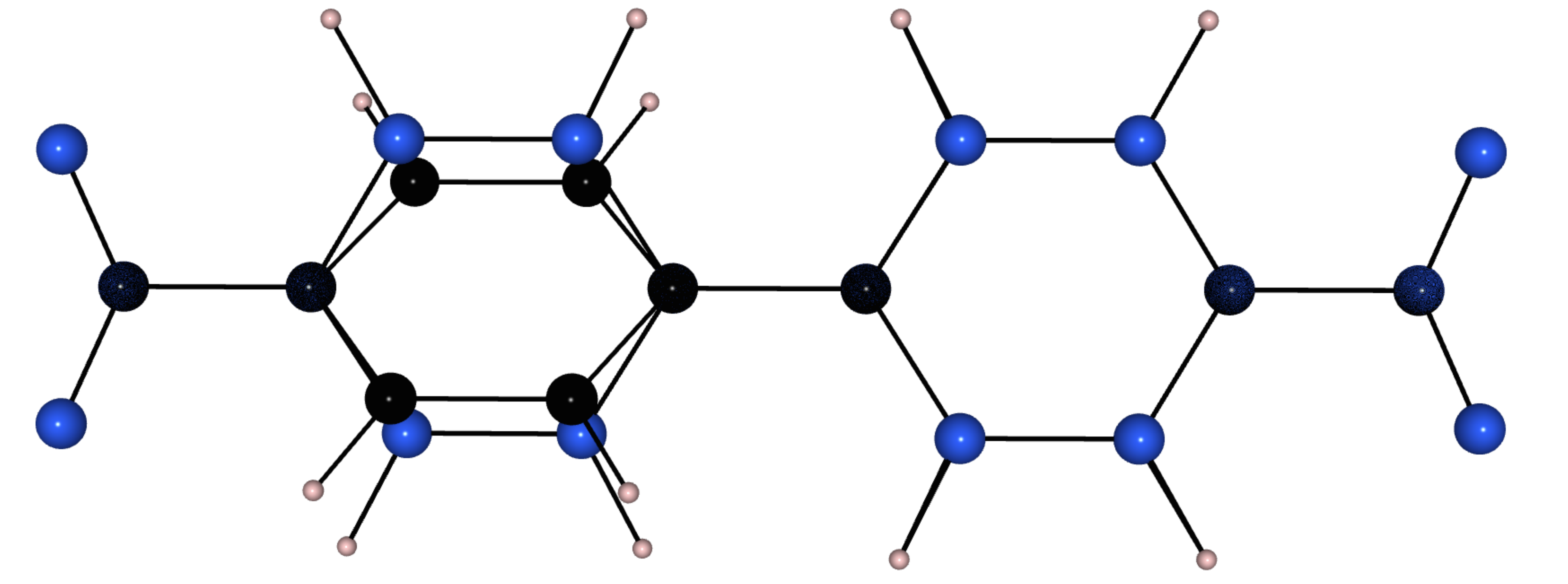}
\caption{Overlaid planar (blue) and twisted (black) 4,4'-BPDC ligands showing the change in geometry of the ligand after following the imaginary phonon modes in the initial structure of IRMOF-10.}
\label{twistligand}
\end{figure}

Prior to the calculation of thermodynamic properties, several observations regarding the vibrational stability of IRMOF-10 and MIL-125 were made during optimisation. Firstly, for IRMOF-10 we initially calculated a significant number of imaginary vibrational modes. To relax the structure into the ground-state and remove all sources of instability, all imaginary modes were simultaneously relaxed following an initial displacement along the corresponding phonon eigenvectors. The final structure was re-converged and no imaginary modes were found. We attribute the initial structural instability to the BPDC ligand; following optimisation we observed a rotation about the central C-C bond connecting the two aromatic rings. The final torsion across this bond was 30-31$^\circ$ between rings (Figure \ref{twistligand}). We propose that the planar experimental configuration may be a thermally averaged structure, and that the true ground-state actually involves twisted ligands. For IRMOF-10 we calculate the structure with twisted ligands to be 0.275 eV (0.046 eV per ligand) more stable than the planar structure. UiO-67 is formed of the same BPDC ligands, which are experimentally characterised as twisted with near identical angles to those in IRMOF-10 between torsion planes. Electronic structure calculations were recently reported by Hemelsoet \textit{et al.}, highlighting the flexibility of the BPDC ligand in UiO-67. This study reported the difference in relative occurrence in torsion angles between the aromatic rings between 0 -- 90$^{\circ}$ during a molecular dynamics simulation. As IRMOF-10 is formed of weaker intermolecular interactions, an increased flexibility of torsion angles would be expected.\cite{van2016vibrational} Furthermore, we calculate the BPDC$^{2-}$ ligand in the gas phase to possess a torsion angle of approximately 33$^{\circ}$ in its ground-state configuration (further details in the SI).

Similar structural instabilities were observed for MIL-125, which is reported as belonging to the tetragonal space group number 139. Initially we obtained 17 associated imaginary modes, and re-optimised to produce a structure with two imaginary modes. This structure was found to have a broken symmetry with the hydrogen on the hydroxyl groups flipped into the pore of the MOF rather than being held in the pore windows. The remaining two imaginary modes could not be removed with further optimisation, and the calculations became too expensive to continue. However the lower symmetry structure was 0.407 eV/primitive unit cell more stable, with no external pressure on the cell.   

The selected MOFs studied here were chosen to ensure a variety of different topologies and bonding interactions were present, thus testing the broad applicability of the forcefield. The metal cations and ligands comprising these MOFs are given, along with the experimentally determined space groups, in Table \ref{MOFparamset}.

\begin{table*}[h]
\caption{MOFs modelled with the \textsc{VMOF} forcefield in this work. Given are the metal cations with the corresponding formal oxidation state and the ligands comprising the MOFs, along with the reported space group number and name.}
\centering
\begin{tabular}{c c c c c} 
\hline
Name  &  Metal   & Ligand  &  Space group   \\
\hline
\\
MOF-5 (IRMOF-1)  & Zn$^{2+}$  &1,4-benzene dicarboxylate  &   225 (Fm$\bar{3}$m) \\
 \\
IRMOF-10 & Zn$^{2+}$ & 4,4'-biphenyl dicarboxylate &  225 (Fm$\bar{3}$m)    \\
\\
 MOF-650 & Zn$^{2+}$  &  2,6-azulenedicarboxylate    &  225 (Fm$\bar{3}$m)   \\
 \\
 UiO-66 & Zr$^{4+}$  & 1,4-benzene dicarboxylate    &  225 (Fm$\bar{3}$m)   \\
 \\
 UiO-67 & Zr$^{4+}$  & 4,4'-biphenyl dicarboxylate  &  225 (Fm$\bar{3}$m) \\
 \\
   MIL-125 & Ti$^{4+}$  &  1,4-benzene dicarboxylate   & 139 (I4/mmm)\\
   \\
 MOF-74 &  Zn$^{2+}$ &   2,5-dihydroxyterephthalic acid    & 2 (P$\bar{1}$)\\
\\
NOTT-300 &  Al$^{3+}$ &  3,3',5,5'-biphenyltetracarboxylic acid    &  98 (I4$_{1}$22)\\
\hline
\end{tabular}
\label{MOFparamset}
\end{table*}

A comparison of optimised unit cell parameters between DFT and forcefield methods are given in Table \ref{MOF}. All structures are reproduced by the forcefield with low errors on the unit cell parameters, thus demonstrating the accuracy of the forcefield and its ability to reproduce different structural features of MOFs, despite the simplicity of its derivation.

\begin{table*}[h]
\caption{Comparison of unit cell parameters (\textit{a}, \textit{b}, \textit{c} with all values in $\mathrm{\AA}$) from the \textsc{VMOF} forcefield with those calculated using DFT (PBEsol functional); percentage errors of FF values compared to DFT are given in brackets.}
\centering
\begin{tabular}{c c c} 
\hline
Name  &  DFT unit cell parameters & \textsc{VMOF} unit cell parameters \\
\hline
\\
MOF-5 (IRMOF-1)(Zn$^{2+}$)    &  25.894,  25.894,  25.894    &  25.935 (0.16),  25.935 (0.16),  25.935 (0.16)    \\
 \\
IRMOF-10 (Zn$^{2+}$)  & 34.385,  34.385,  34.385  &   34.417 (0.09), 34.417 (0.09), 34.417 (0.09)     \\
\\
 MOF-650 (Zn$^{2+}$)   & 30.695, 30.695, 30.695   & 30.766 (0.23), 30.766 (0.23), 30.766 (0.23)        \\
 \\
               MOF-74 (Zn$^{2+}$) & 6.740,  15.142, 15.142,  &  6.764 (0.35), 15.031 (0.73), 15.031 (0.73)     \\
              \\
 UiO-66 (Zr$^{4+}$)   &   20.798, 20.798, 20.798   &  20.909 (0.53), 20.909 (0.53), 20.909 (0.53)    \\
 \\
 UiO-67 (Zr$^{4+}$) & 27.094, 27.094, 27.094  &  26.878 (0.80),  26.878 (0.80),  26.878 (0.80)  \\
 \\
   MIL-125 (Ti$^{4+}$) & 18.852, 18.843, 17.921 &   18.859 (0.01), 18.859 (0.08), 18.043 (0.68)  \\
   \\
                 NOTT-300 (Al$^{3+}$) &   14.836, 14.836,  11.871   &   14.862 (0.18),  14.862 (0.18), 11.500 (3.23)   \\
\hline
\end{tabular}
\label{MOF}
\end{table*}

\subsection{Mechanical properties}

Bulk moduli have been calculated with DFT and \textsc{VMOF} to compare the mechanical strength of the materials predicted with the two approaches (Table \ref{bulk moduli}).

\begin{table*}[h]
\caption{Comparison of bulk moduli (GPa) obtained with the \textsc{VMOF} forcefield with those calculated with DFT. DFT bulk moduli were calculated with the PBEsol functional with D3 dispersion correction, by fitting calculated energy/volume curves to a Birch-Murnaghan equation of state with +/- 3 $\%$ sampling away from the equilibrium volume.}
\centering
\begin{tabular}{c c c c} 
\hline
Name (metal) &  DFT (GPa) & \textsc{VMOF} (GPa)   \\
\hline
\\
MOF-5/IRMOF-1 (Zn$^{2+}$) &  16.9  &    8.8 \\
 \\
IRMOF-10 (Zn$^{2+}$)& 8.6  &  5.1     \\
\\
 MOF-650 (Zn$^{2+}$)& 12.5   &    6.8     \\
 \\
 UiO-66 (Zr$^{4+}$)&  40.4  &  19.0   \\
 \\
 UiO-67 (Zr$^{4+}$)& 21.9   &  11.7 \\
 \\
   MIL-125 (Ti$^{4+}$) &   25.1    &18.5  \\
   \\
 MOF-74 (Zn$^{2+}$) & 28.1    & 14.9  \\
\\
NOTT-300 (Al$^{3+}$) & 47.8 & 25.2  \\
\hline
\end{tabular}
\label{bulk moduli}
\end{table*}

The forcefield predicts smaller bulk moduli than DFT. We do, however, highlight that experiment often finds MOFs to have a softer bulk modulus than those predicted with electronic structure methods. Yot \textit{et al.} reported experimental bulk moduli, as measured using high pressure XRD methods, for UiO-66 and MIL-125 to be 17.0 and 10.0 GPa, respectively, which are closer to the forcefield values than those calculated by DFT.\cite{yot2016exploration} The trend in mechanical strength for each framework is the same between the two methods, with increasing in mechanical strength from IRMOF-10 $<$ MOF-74 $<$ MOF-650 $<$ MOF-5 $<$ UiO-67 $<$ MIL-125 $<$ UiO-66 $<$ NOTT-300.  

Higher charged cations form primarily ionic interactions between metal and ligand. It can therefore be rationalised that UiO-66, UiO-67 and MIL-125 would possess stiffer bulk moduli (greater resistance to compression) than the Zn-isorecticular MOFs, which possess large internal voids and weaker van der Waals interactions between Zn$^{2+}$ and organic ligands.

\subsection{Phonon properties}

\subsubsection{IR spectra. }

\begin{figure*}
\centering
\includegraphics[width=12cm]{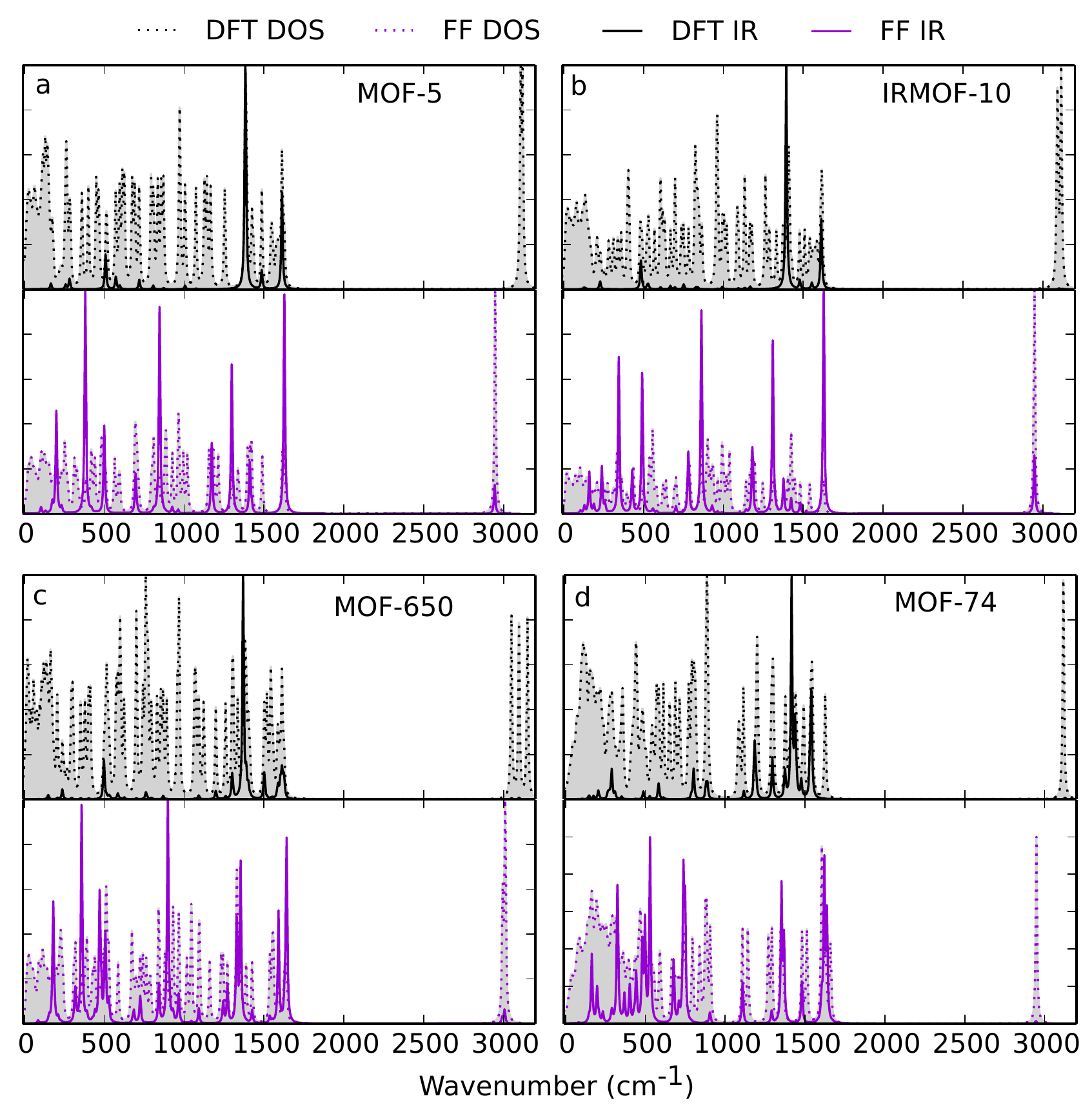}
\caption{Overlaid IR spectra and phonon density of states (DOS) calculated with DFT (top, black) and FF (bottom, purple) methods, plotted between 0--3200 cm$^{-1}$ for MOF-5 (a), IRMOF-10 (b), MOF-650 (c) and MOF-74 (d). All spectra are normalised to lie between 0 and 1 and area the under DOS is shaded (grey) for clarity.}
\label{IR_1}
\end{figure*}

\begin{figure*}
\centering
\includegraphics[width=12cm]{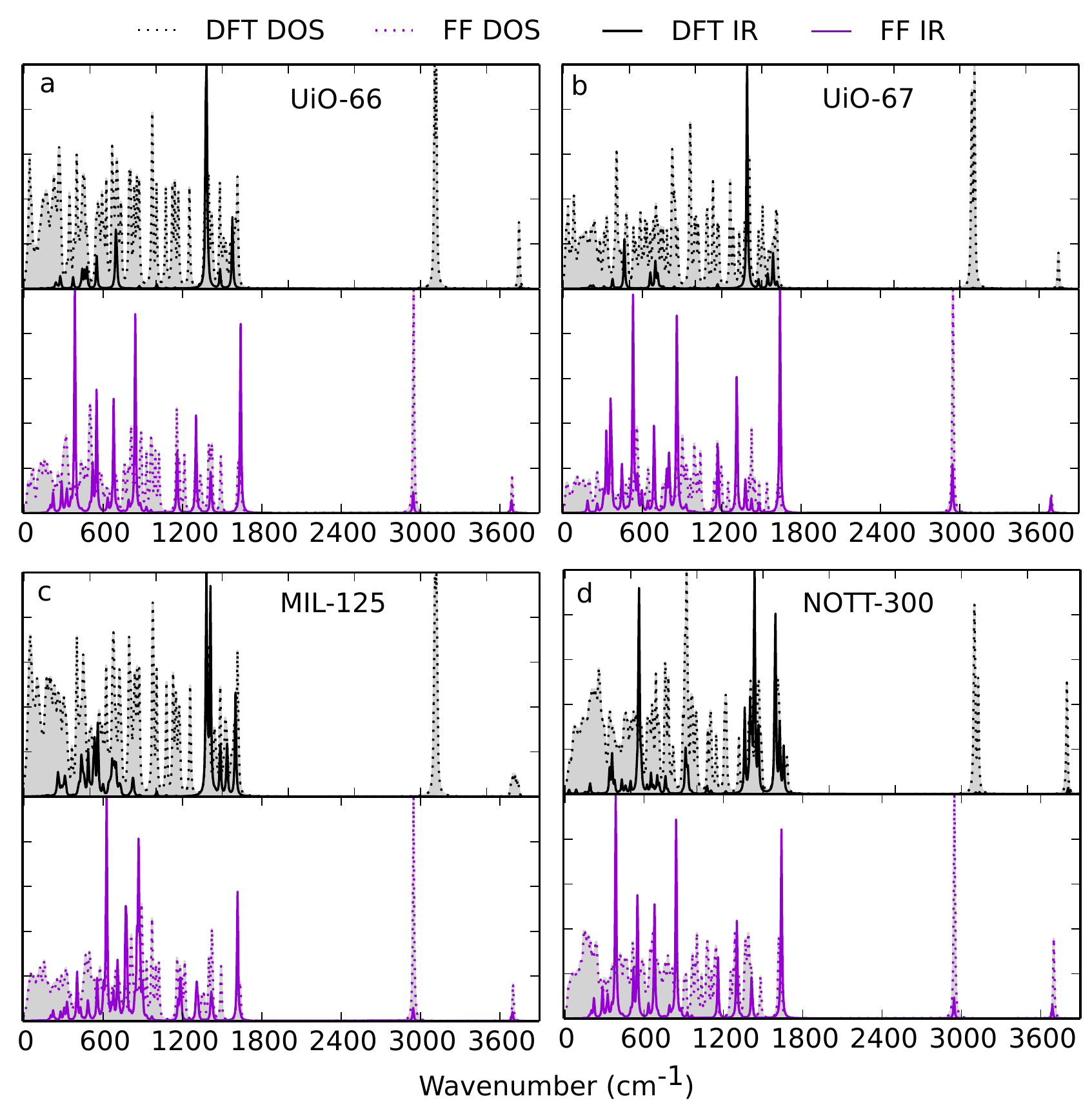}
\caption{Overlaid IR spectra and phonon density of states (DOS) calculated with DFT (top, black) and FF (bottom, purple) methods, plotted between 0--3900 cm$^{-1}$ for UiO-66 (a), UiO-67 (b), MIL-125 (c) and NOTT-300 (d). All spectra are normalised to lie between 0 and 1 and area under the DOS is shaded (grey) for clarity.}
\label{IR_2}
\end{figure*}

The first approach to assessing the ability of a forcefield to reproduce accurate vibrational properties is to calculate the phonon density of states and the associated IR spectra (weighted by the mode intensities) to ensure the fingerprint of vibrational modes is similar between DFT and forcefield methods. Good agreement is observed between vibrational IR spectra and DOS between DFT (Figure \ref{IR_1}) and the forcefield (Figure \ref{IR_2}). Comparison of the plots highlights the stability of the modelled MOFs with both methods, but also the small deviation between the two sets of calculated DOS and IR spectra. The biggest discrepancy between \textsc{VMOF} and DFT IR spectra is in the fingerprint (lower frequency region). This is due to the metal-oxygen bond stretching modes, and since IR activity $\propto$ charge $\times$ displacement, the discrepancy is primarily due to the use of formal charges of the metal ions. Importantly, we highlight that the DOS spectra between DFT and forcefield remain comparable, which suggests the use of a formal charge model has had little effect on the forces. 

\begin{figure*}
\centering
\includegraphics[width=12cm]{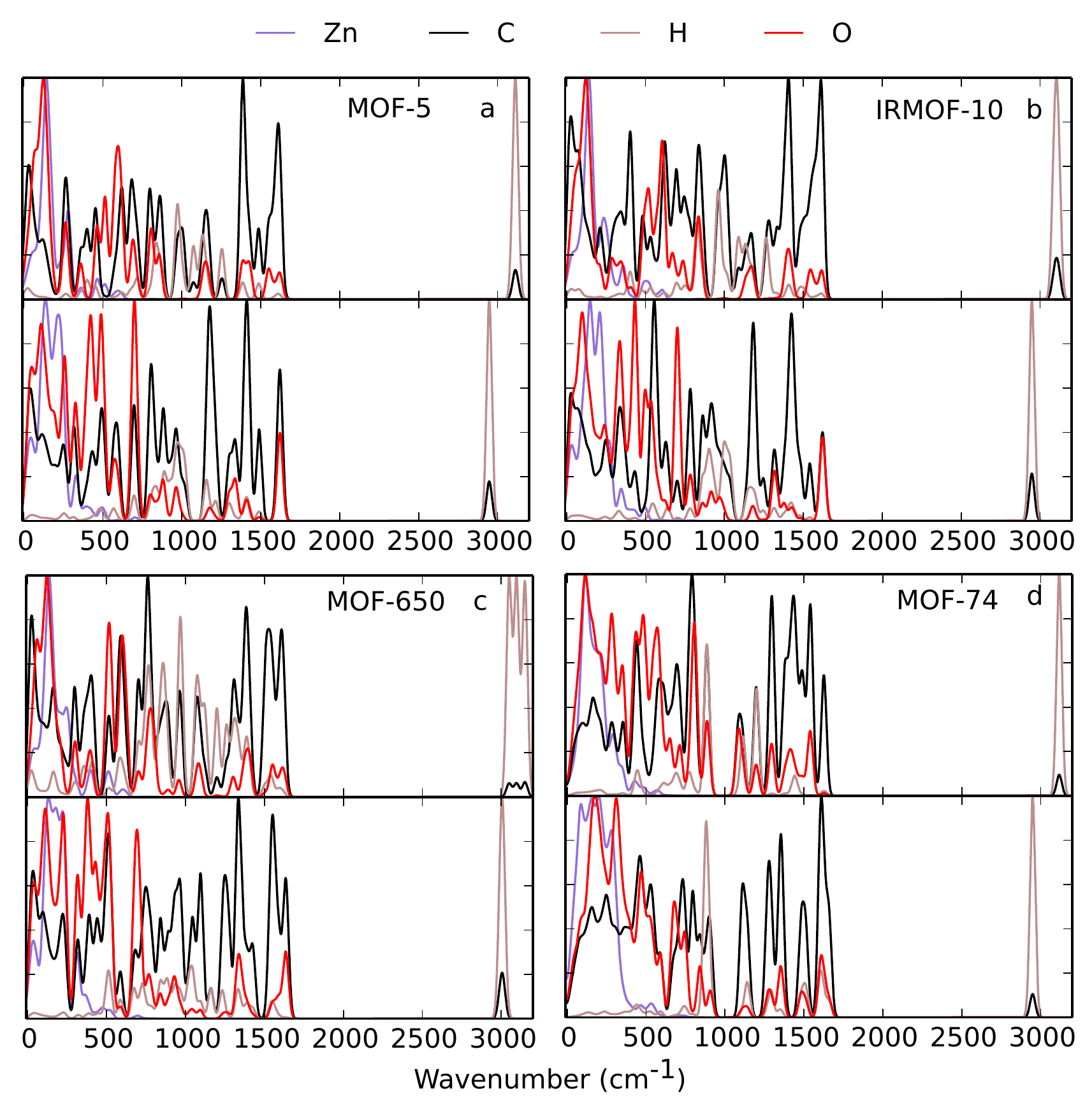}
\caption{Partial phonon density of states (PDOS) projected according to the elemental contribution calculated with DFT (top) and FF (bottom) methods, plotted between 0--3200 cm$^{-1}$ for MOF-5 (a), IRMOF-10 (b), MOF-650 (c) and MOF-74 (d). All spectra are normalised to lie between 0 and 1.}
\label{PDOS_1}
\end{figure*}

\begin{figure*}
\centering
\includegraphics[width=12cm]{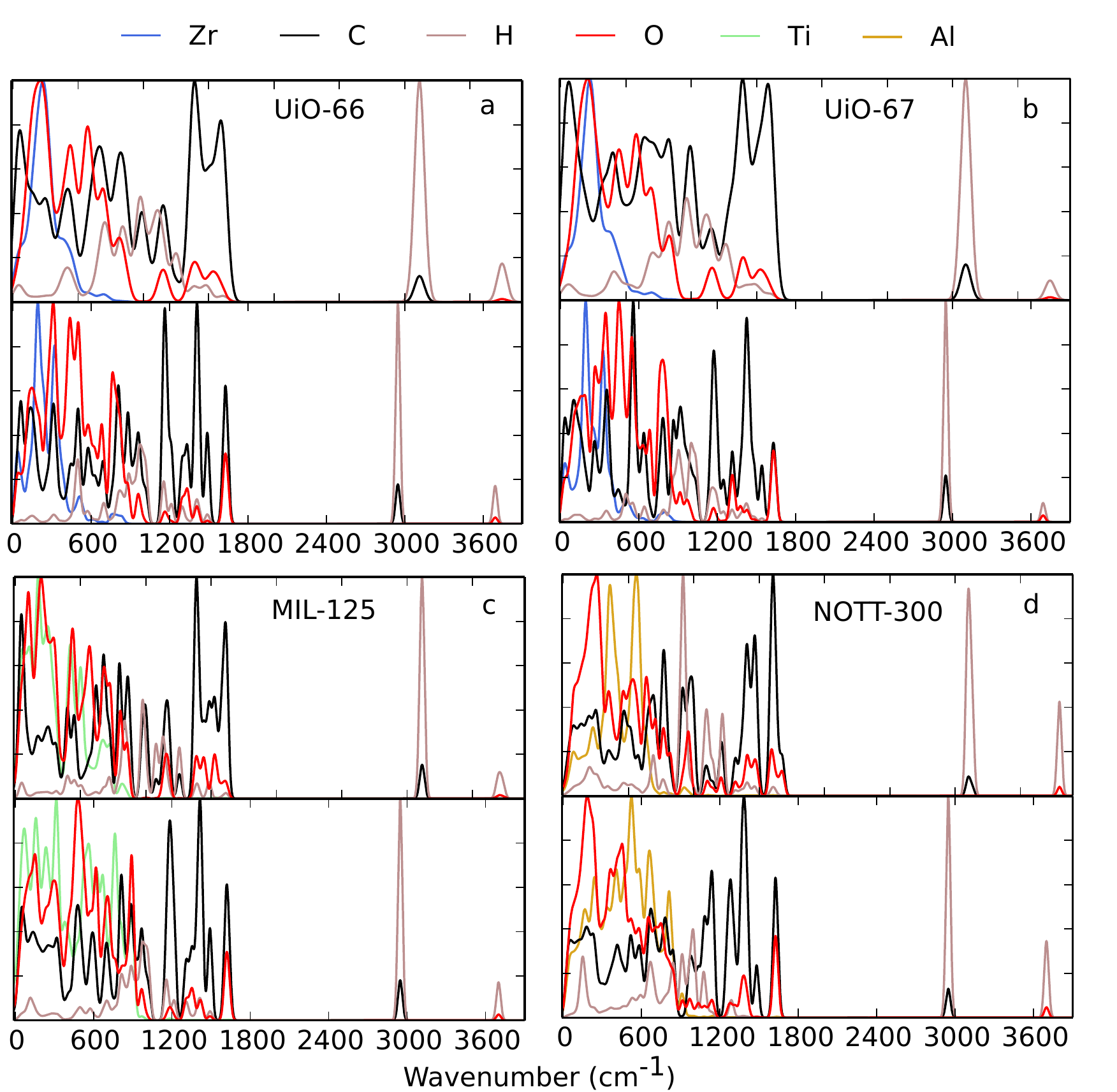}
\caption{Partial phonon density of states (PDOS) decomposed according to the elemental contribution calculated with DFT (top) and FF (bottom) methods, plotted between 0--3900 cm$^{-1}$ for UiO-66 (a), UiO-67 (b), MIL-125 (c) and NOTT-300 (d). All spectra are normalised to lie between 0 and 1.}
\label{PDOS_2}
\end{figure*}

The decomposition of the DOS into elemental contributions is shown in the partial DOS plots (Figures \ref{PDOS_1} and \ref{PDOS_2}). Several features are evident in both spectra calculated with DFT and FF methods that are chemically well established. Firstly, the range of modes involving the metal cations all occur at low frequencies ($<$ 500 cm$^{-1}$). Also observed in this region for all MOFs is a small contribution from C, H and O from the rocking motions of the ligands. At finite temperature it is these low frequency modes that are populated, and therefore control the MOF dynamics (\textit{e.g.} the shape of the thermal ellipsoid). Therefore, the motions of the MOF will occur primarily at the metal nodes, as well as subtle rotations at the MOF-ligand connections. In the mid-frequency range, between 1300 -- 1500 cm$^{-1}$, the high contribution of C and O to the density of states is due to motions of the asymmetric and symmetric C$_{carb}$-O$_{carb}$ stretches of the carboxylate groups in the MOFs. Finally, modes above 3000 cm$^{-1}$ are associated with C-H and O-H stretches at the ligand and within some metal nodes, respectively. The important conclusion from the density of states plots is that we see good agreement between DFT and forcefield methods, suggesting the vibrational properties of the MOFs are well reproduced by \textsc{VMOF}.

It is common to characterise the vibrational properties of a material by assigning specific IR frequencies. However, we demonstrate that a vast amount of vibrational information is not accounted for by doing this for MOFs. The comparison of DOS and IR spectra show the difference in detail and highlight the importance of considering vibrational modes that are not IR active when parametrising a forcefield. The DOS also shows the significant number of soft vibrational modes that MOFs possess, which give rise to structural instability with temperature and pressure. 
We note a shift of the frequencies of the C-H stretch between the two methods. As the C-H stretch occurs as one of the highest frequency modes, it is contributing the most to the zero-point vibrational energy and is likely to be the biggest contribution to the C-H stretch calculated error between methods. The reliability of the forcefield is not likely to be affected by the disagreement in zero point energy between the methods for the study of complex processes such as phase changes. It is unclear if it is DFT or the FF model that contains the greatest error on the C-H stretch. The forcefield parameters for the C-H interaction remain unchanged from the CHARMM forcefield and therefore have not been derived specifically for BDC incorporated into a MOF. On the other hand, a scaling factor is often used on the vibrational frequencies in DFT calculations that would have the greatest effect on the stretching mode of the C-H bond. Such scaling factors can correct for anharmonicity, while forcefield parameters can be derived explicitly to reproduce experimental anharmonic frequencies.

\subsubsection{Helmholtz free energy. }

Reproducing accurate relative free energies of a system with temperature is crucial for the prediction of thermodynamic processes such as phase changes and reaction energies for the formation of defects within frameworks, for example in our previous work investigating the ``missing linker phenomenon'' in UiO-66.\cite{bristow2016free}

\begin{figure*}
\centering
\includegraphics[width=12.0cm]{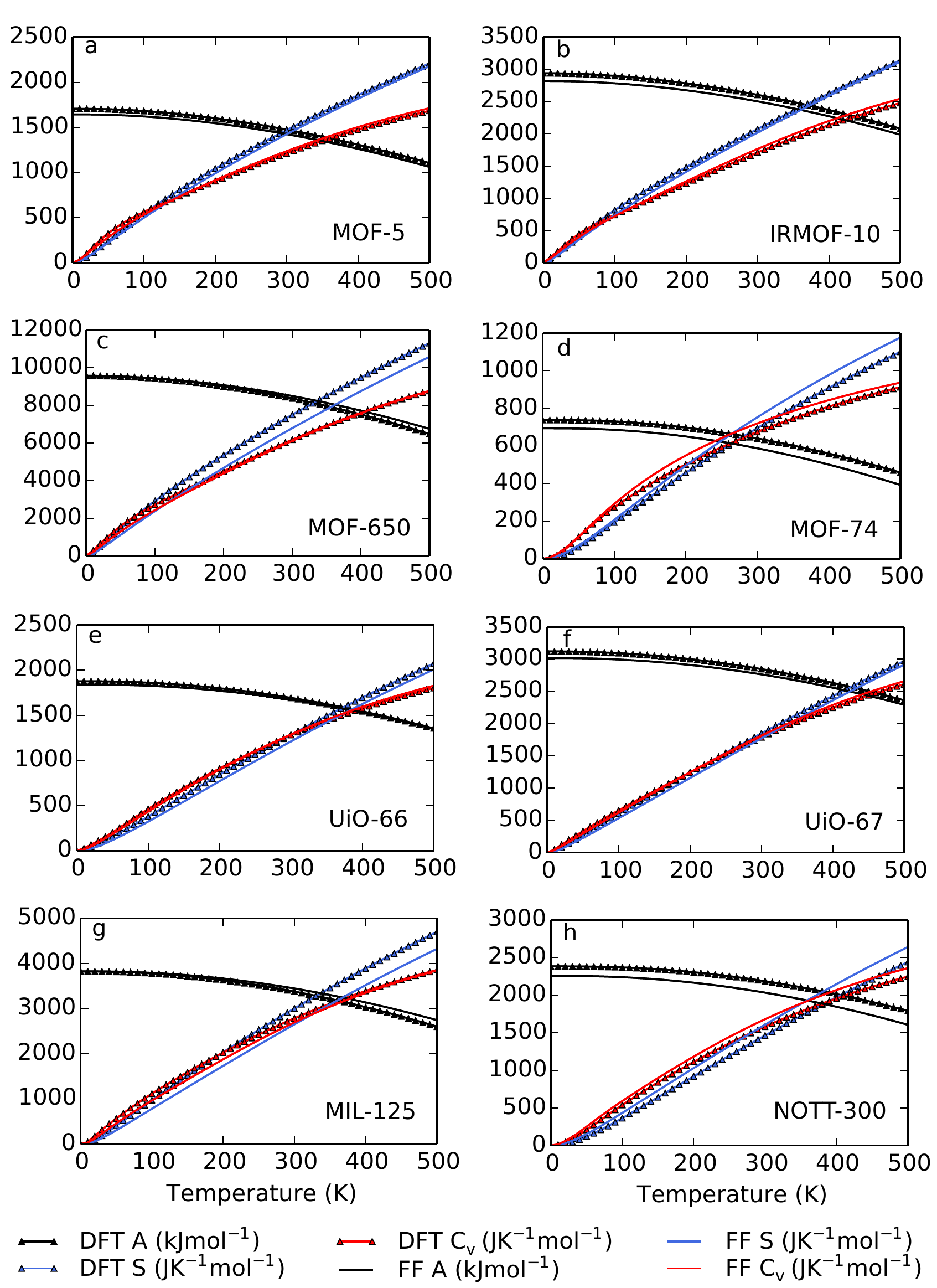}
\caption{Comparison of calculated Helmholtz free energy (black), vibrational entropy (blue) and constant volume heat capacity (C$_{v}$) (red) of (a) MOF-5, (b) IRMOF-10, (c) MOF-650, (d) MOF-74, (e) UiO-66, (f) UiO-67, (g) MIL-125 and (h) NOTT-300 between DFT (symbol) and FF (line) methods.}
\label{helmholtz}
\end{figure*}

The constant volume (Helmholtz) free energy, vibrational entropy and constant-volume heat capacities as a function of temperature are plotted for all MOFs considered with both first principles and forcefield methods (Figure \ref{helmholtz}). The forcefield is shown to reproduce the calculated thermodynamic properties from DFT very well. Little deviation across all structures is observed, further supporting that the forcefield can accurately reproduce the vibrational properties of the subset of MOFs studied.

\subsubsection{Quasi-harmonic approximation properties. }

Fitting forcefield models to ground state properties of the equilibrium structure does not account for the variation in volume and unit cell shape with temperature. As a method for extending the harmonic model of vibrations, the quasi-harmonic approximation (QHA) allows anharmonicity associated with volume change (thermal expansion) to be considered when calculating structural and thermodynamic properties.

Whilst conducting tests for the quasi-harmonic approximation we noted a particular sensitivity of the fitted forcefield parameters to the free energy equation of state with framework compressions. Consequently, calculated properties varied depending on the volume sampled for the expansions and contractions away from the local minimum structure. For MOF-5 the initial volume sampled was +/-3$\%$ away from the energy minimum in 0.05$\%$ steps. We observed, with both DFT and forcefield methods, that beyond approximately 2$\%$ compression multiple imaginary vibrational modes emerged. When following these imaginary modes, a subtle twist at the carboxylate head relative to the benzene ring (approximately 3 -- 5$^\circ$) was observed. The same observation was made for IRMOF-10, suggesting that with compression ``softer'' MOFs undergo this subtle rotation of the benzene rings leaving a non-planar torsion between the ring and carboxylate heads of the ligand. We note that we could remove imaginary modes for all structures during compression with the forcefield, but the same process with electronic structure methods became too expensive for IRMOF-10, which possessed 2 imaginary modes at the highest compression of 3$\%$. 

\begin{figure}
\centering
\includegraphics[width=7.0cm]{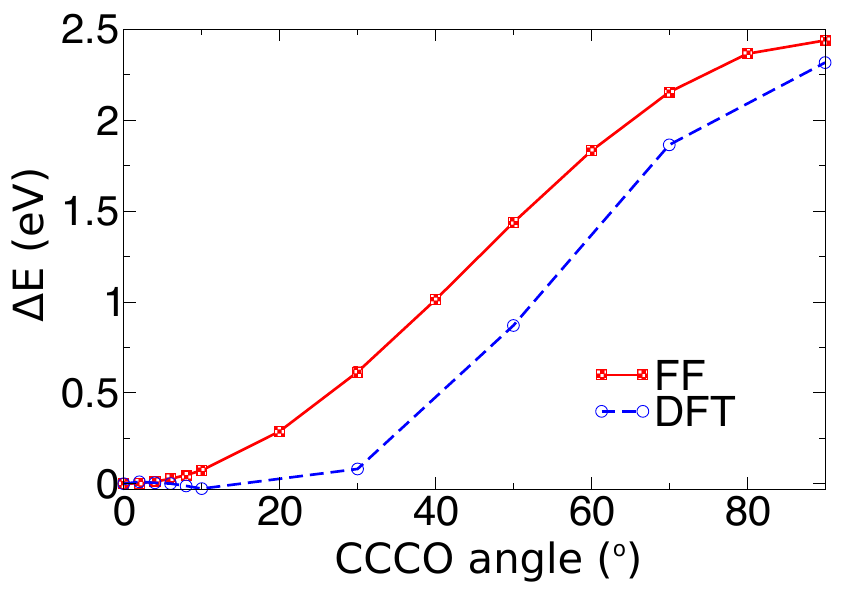}
\caption{Energy cost per ligand when rotating all C-C-C-O torsions in the ligands by successive angle increments away from the initial planar 0$^\circ$ structure. Relative energies are reported for both DFT and \textsc{VMOF}.}
\label{rotationmof5}
\end{figure}

To investigate this phenomenon further we modelled the same rotation of BDC in MOF-5 away from the initial planar 0$^\circ$ structure. Calculations were conducted on the periodic systems with the \textsc{VMOF} forcefield. To model the same rotation with DFT, we cut a representative cluster with the chemical formula Zn$_{4}$O(BDC-H$_{2}$). The PBE0 functional was used in the NWChem program with the Dunning correlation consistent cc-pVTZ basis sets.\cite{valiev2010nwchem,dunning1989gaussian,woon1993gaussian} 
 
The rotation leaves the carboxylate heads in the same position in a direct bonding interaction with the metal centres, and only moves the benzene ring; following the rotation, there is a non-planar torsion between ring and carboxylate head (see SI). Rotating the ligand in this manner in MOF-5 is shown to cost little energy with both DFT and \textsc{VMOF} up to 10$^\circ$ (Figure \ref{rotationmof5}). The potential energy surface is shallow, with a 5$^{\circ}$ change being comparable to $k_B$T. Therefore it can be expected that the group will be rotationally active at room temperature. We highlight the similarity in the calculated trend between DFT and \textsc{VMOF} for the modelled rotation, supporting our observations of the instability of the planar structure with compression with DFT.  

\begin{figure*}
\centering
\includegraphics[width=12cm]{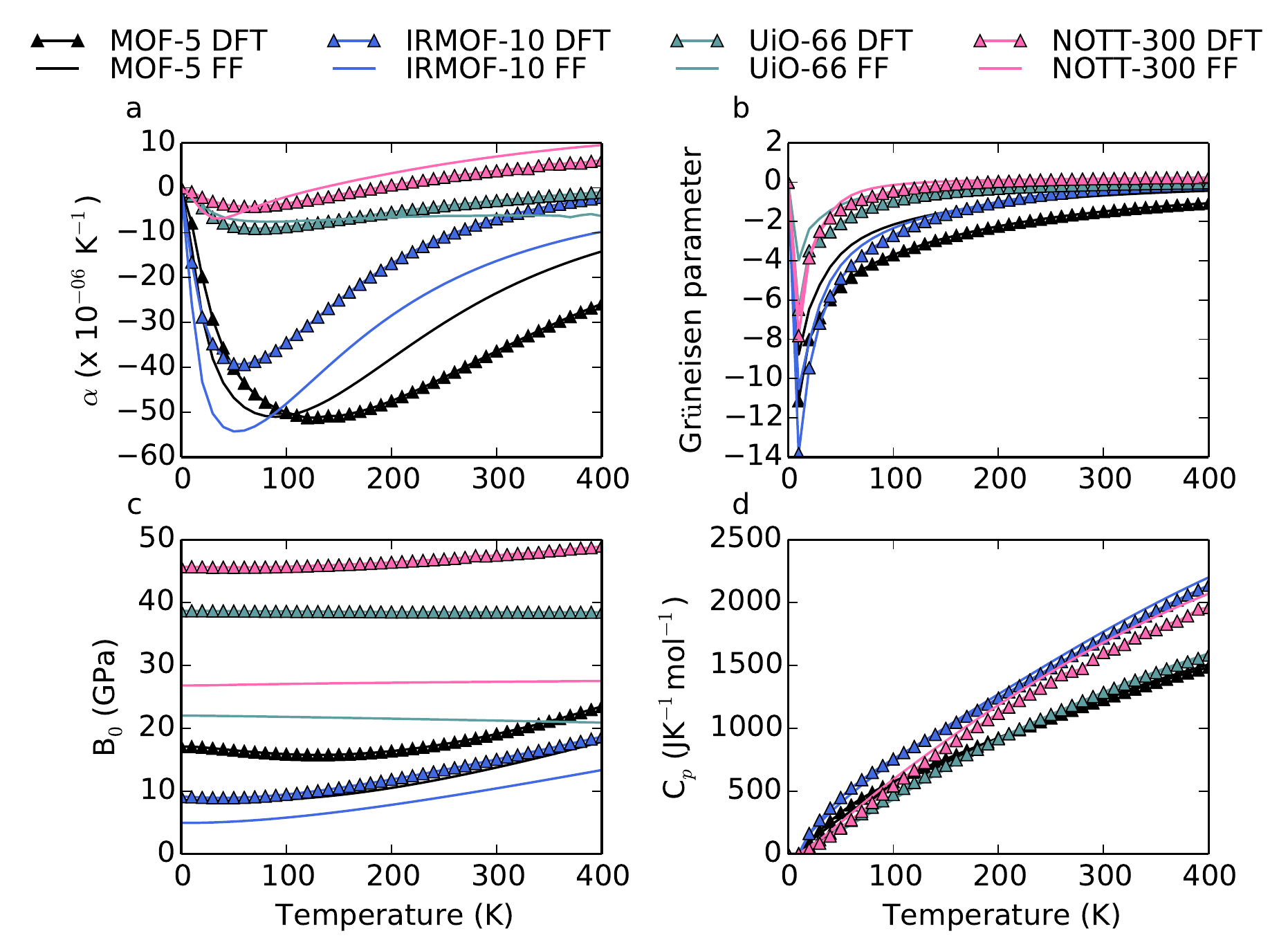}
\caption{Comparison of structural and thermodynamic properties as a function of temperature calculated with the quasi harmonic-approximation for MOF-5, IRMOF-10, UiO-66 and NOTT-300 using DFT and the \textsc{VMOF} forcefield (FF).
(a) linear thermal expansion coefficient
(b) Gr$\mathrm{\ddot{u}}$neisen parameter
(c) bulk modulus
(d) heat capacity at constant pressure.}
\label{qhaall}
\end{figure*}

\begin{figure*}
\centering
\includegraphics[width=12cm]{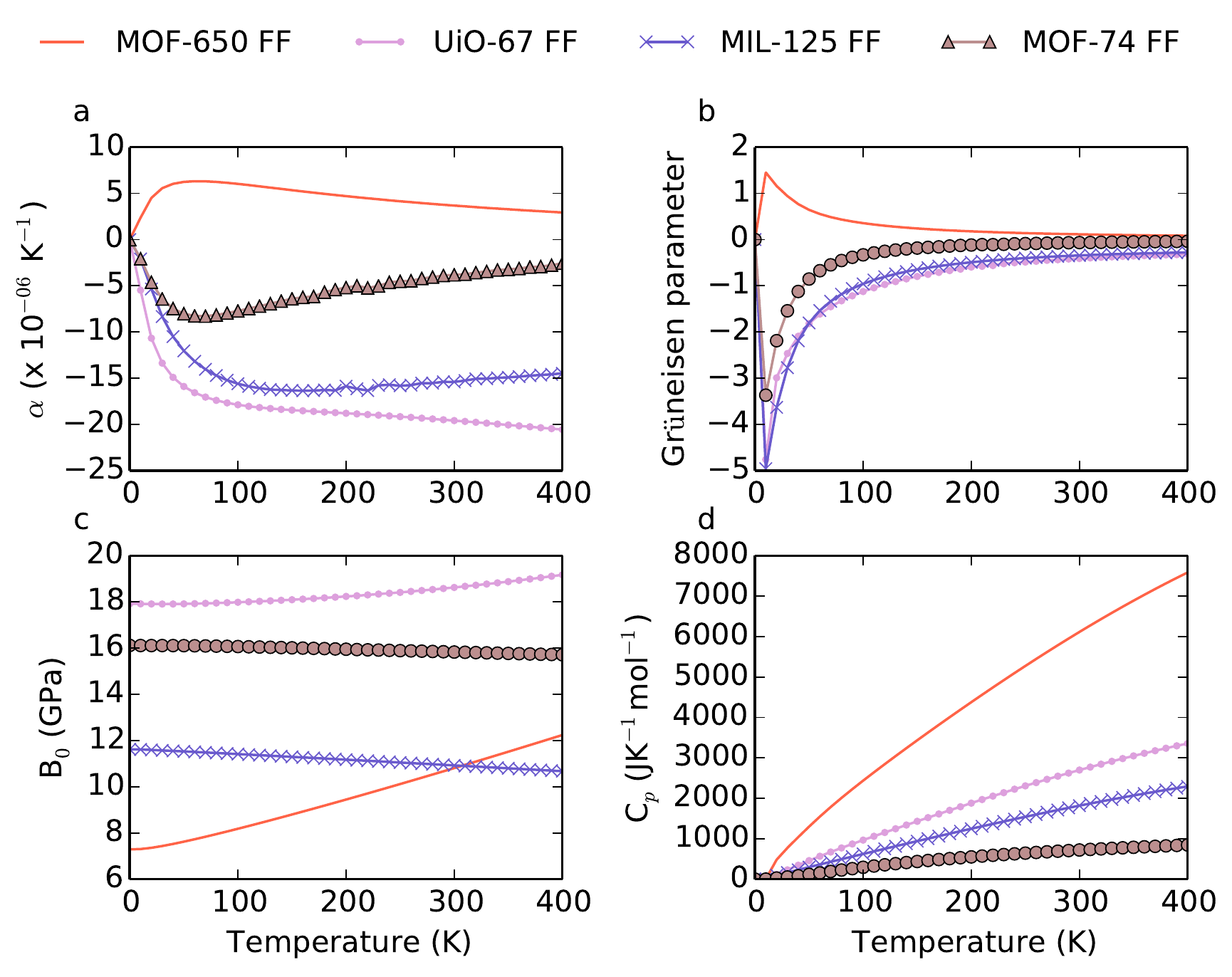}
\caption{Comparison of structural and thermodynamic properties as a function of temperature calculated with the quasi-harmonic approximation for MOF-650, MOF-74, MIL-125 and UiO-67 using DFT and the \textsc{VMOF} forcefield (FF).
(a) linear thermal expansion coefficient  
(b) Gr$\mathrm{\ddot{u}}$neisen parameter
(c) bulk modulus
(d) heat capacity at constant pressure.}
\label{qhaall2}
\end{figure*}

Once the physical significance of structural distortion with compression was investigated, we were able to rationalise trends observed from quasi-harmonic calculations. Due to the expense of the method with first principles calculations, we report a comparison of properties for DFT and \textsc{VMOF} for only four structures; MOF-5, IRMOF-10, UiO-66 and NOTT-300 (Figure \ref{qhaall}), while quasi-harmonic calculations are also performed for MIL-125, UiO-67, MOF-650 and MOF-74 with the forcefield (Figure \ref{qhaall2}).

The use of the quasi-harmonic approximation allows the calculation of a more extensive range of properties, such as the Gr$\mathrm{\ddot{u}}$neisen parameter, bulk modulus, heat capacity and thermal expansion coefficient, as a function of temperature. For all structures we found that the QHA properties that showed the greatest sensitivity were the bulk modulus and thermal expansion. These properties are derived directly and indirectly from the curvature of the potential energy surface, respectively, and therefore exaggerate the difference between the methods. We highlight that \textsc{VMOF}, which was not specifically parameterised to reproduce negative thermal expansion of MOFs, manages to exhibit this phenomenon and accurately in comparison to DFT for MOF-5, IRMOF-10, UiO-66 and NOTT-300.

Whilst fitting an equation of state for IRMOF-10, we observed a sensitivity of the calculated properties with DFT to the extent of compression considered. Properties such as the bulk modulus, Gr$\mathrm{\ddot{u}}$neisen parameter and thermal expansion were calculated to vary significantly when including high-pressure points (\textit{i.e.} large compressions). We also observed a significant shift in minimum of Helmholtz energy with change in volume at the defined temperatures. Such a shift and variation in calculated properties suggests the zero point energy contribution to have a large effect on the structure.  

Several interesting features more specific to each structure are observed from the QHA calculations (Figures \ref{qhaall} and \ref{qhaall2}). Firstly, a comparison of thermal expansion coefficients show the vast difference in structural changes with temperature. The ``softer'' Zn-isorecticular MOFs, such as MOF-5 and IRMOF-10 with both DFT and FF methods, are calculated to possess the largest negative thermal expansion coefficients at low temperature, suggesting the extent of thermal expansion reflects the mechanical strength of the materials. Indeed, the ``hardest'' MOFs, UiO-66 and NOTT-300, show little variation in thermodynamic properties with temperature.
The temperature dependent bulk moduli are shown to differ between DFT and forcefield methods (Figure \ref{qhaall}), following the calculated trend in static bulk moduli. Specifically, \textsc{VMOF} yields softer mechanical properties at finite temperatures than electronic structure methods. We note that the trend and shape of each profile with temperature is reproduced by the forcefield, and that the temperature dependent Gr$\mathrm{\ddot{u}}$neisen parameter and constant pressure heat capacity at constant pressure are reproduced accurately and appear unaffected by the discrepency in bulk moduli. Due to the accuracy of the quasi-harmonic approximation being restricted to $\frac{1}{2}$ -- $\frac{2}{3}$ the melting point temperature (depending on the material), we cannot model the full behaviour of the heat capacity at high temperatures. 
The temperature dependence of the thermodynamic properties of MOF-650 appears to show a significantly different trend to all other structures, which all show similar behaviour. MOF-650 has a large internal void with cell parameters exceeding 30 $\mathrm{\AA}$. The azulene ligand comprising MOF-650 is also rigid and would allow little structural flexibility when compared to IRMOF-10, which is of similar size. It is likely to be these two factors that result in positive thermal expansion and Gr$\mathrm{\ddot{u}}$neisen parameter at low temperature.
Finally, the Gr$\mathrm{\ddot{u}}$neisen parameter shows a similar trend for each structure, excluding MOF-650. The increase in Gr$\mathrm{\ddot{u}}$neisen parameter with temperature reflects the increase in mechanical strength following contraction of the cell parameters.

\section{Conclusions}

A new transferable forcefield for metal-organic frameworks named \textsc{VMOF} has been parameterised to reproduce accurate lattice dynamics and phonon properties. Such a forcefield contributes greatly to the current extensive field of MOF forcefields as it is unique in the number of thermodynamic properties that can be accurately determined in a rapid and transferable manner. For an initial training set of MOFs including MOF-5, IRMOF-10, UiO-66, UiO-67, NOTT-300, MIL-125, MOF-650 and MOF-74 we calculate numerous thermodynamic properties including bulk modulus, free energies and constant volume heat capacities. We further conduct quasi-harmonic calculations and find excellent agreement in thermal expansion, bulk moduli, Gr$\mathrm{\ddot{u}}$neisen parameter and heat capacity with temperature between DFT and the newly parameterised forcefield. This now opens the way for the future high-throughput computational screening of materials vibrational properties for a wide range of MOFs.

\section{Acknowledgments} 
J.K.B is funded by the EPSRC (Grant No. EP/G03768X/1). J.D.G thanks the Australian Research Council for funding under the Discovery Programme, as well as the Pawsey Supercomputing Centre and NCI for the provision of computing resources. A.W. acknowledges support from the Royal Society University Research Fellowship scheme. K.L.S. is funded under ERC Starting Grant 277757 and J.M.S is funded under EPSRC grant no. EP/K004956/1. The work benefited from the high performance computing facility, Balena, at the University of Bath, and access to the ARCHER supercomputer through membership of the HPC Materials Chemistry Consortium (EP/L000202).  

\section{Data Access Statement} 
The \textsc{VMOF} forcefield is available as a library file for \textsc{GULP} from \url{https://github.com/WMD-group/VMOF}. 
A set of raw phonon data and program input/output files are available from \url{https://researchdata.ands.org.au}.

\footnotesize{
\bibliography{mofbib} 
\bibliographystyle{rsc} 
}

\end{document}